\documentclass[reprint,groupedaddress,amsmath,amssymb,aps,prx,floatfix,longbibliography]{revtex4-1}

\usepackage[utf8]{inputenc}
\usepackage{amsmath,amssymb,graphicx,mathrsfs,amsfonts,physics, float, placeins}

\usepackage{dsfont}
\usepackage{color}

\begin{document}
\title{Role of Atoms in Atomic Gravitational-Wave Detectors }
\author{Matthew A. Norcia}
\affiliation{JILA, NIST, and University of Colorado, 440 UCB, 
Boulder, CO  80309, USA}
\author{Julia R. K. Cline}
\affiliation{JILA, NIST, and University of Colorado, 440 UCB, 
Boulder, CO  80309, USA}
\author{James K. Thompson}
\affiliation{JILA, NIST, and University of Colorado, 440 UCB, 
Boulder, CO  80309, USA}
\email[]{mattanorcia@gmail.com}

\begin{abstract}
Recently, it has been proposed that space-based atomic sensors may be used to detect gravitational waves.  
These proposals describe the sensors either as clocks or as atom interferometers.  
Here, we seek to explore the fundamental similarities and differences between the two types of proposals. 
We present a framework in which the fundamental mechanism for sensitivity is identical for clock and atom interferometer proposals, with the key difference being whether or not the atoms are tightly confined by an external potential.  
With this interpretation in mind, we propose two major enhancements to detectors using confined atoms, which allow for an enhanced sensitivity analogous to large-momentum-transfer (LMT) used in atom interferometry (though with no transfer of momentum to the atoms), and a way to extend the useful coherence time of the sensor beyond the atom's excited state lifetime.


\end{abstract}
\maketitle

\section{Introduction}
The recent observation of gravitational waves (GWs) \cite{Ligo1, Ligo2} has established gravitational wave detection as an exciting new observational tool for cosmological phenomena.  Terrestrial optical interferometers are sensitive to gravitational radiation at frequencies above roughly 10 Hz \cite{harry2010advanced}.  In order to extend these techniques to lower frequencies where there are expected to be an abundance of signals, space-based optical interferometers such as the Laser Interferometer Space Antenna (LISA) have been proposed, and are currently under technological development \cite{LISA_proposal, LISApathfinder}.  

\begin{figure*}[!htb]
\includegraphics[width=6.75in]{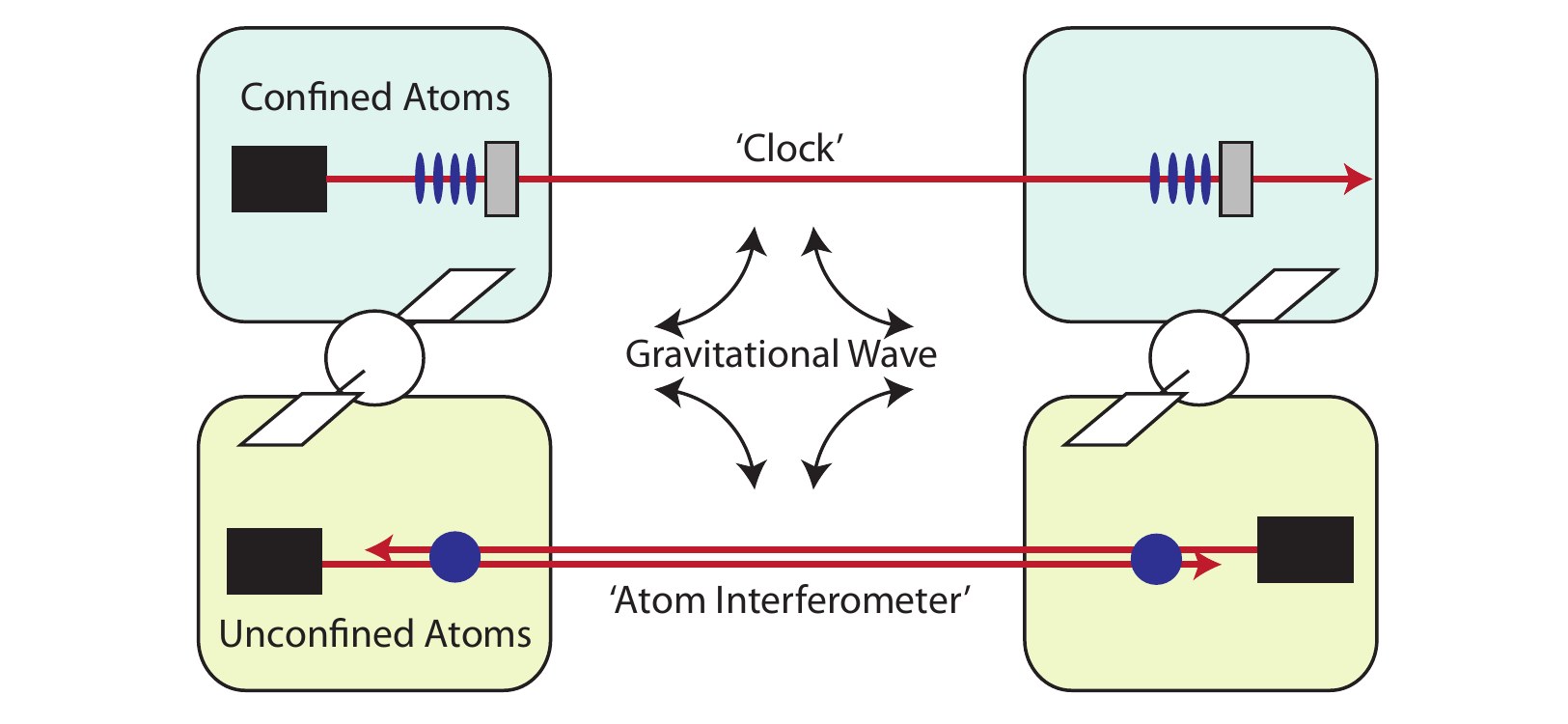}
\caption{Optical lattice clock and atom interferometer detectors for gravitational waves.  Each type of proposal relies on two ensembles of atoms, one in each satellite.  Lasers transmitted between the satellites encode phase shifts due to changing optical path lengths caused by a gravitational wave.  The key difference between clock and atom interferometer proposals is that in clocks, atomic recoils due to the momentum of absorbed/emitted photons are supressed by tightly confining the atoms, while in atom interferometers the atoms are free to recoil.  }
\label{fig:ExpDiag}
\end{figure*}

More recently, space-based gravitational wave detectors based on optical transitions in cold atoms have been proposed as an alternative architecture \cite{2015arXiv150100996L, GravitationalRaman2009, yu2011gravitational, kolkowitz2016gravitational, vutha2015optical, hollberg2017optical, graham2013new, hogan2016atom, graham2016resonant}.  These proposals rely on a combination of optical and atomic coherence to provide sensitivity to gravitational waves. While optical interferometers require three satellites to cancel laser phase noise, these proposals require only two.  
We will focus here on two categories of the atom-based proposals:  those described as ``atom interferometers" (AI) typified by \cite{graham2013new}, and those described as ``optical lattice clocks" \cite{kolkowitz2016gravitational}.  These two types of proposal are illustrated in Fig.~\ref{fig:ExpDiag}.  

AI-type proposals based on two-photon Raman transitions have also been proposed as a means of detecting gravitational waves, both in space-based and ground-based applications \cite{dimopoulos2009gravitational, canuel2014matter, PhysRevD.93.021101}. However, these proposals are not inherently insensitive to laser phase noise, and thus require noise cancellation techniques similar to a purely optical interferometer, such as the use of three interferometers.  AI detectors based on two-photon Raman transitions may prove to be powerful tools for detecting gravitational waves, but here will focus exclusively on proposals that leverage long-lived optical atomic coherence on a single detection baseline with intrinsic insensitivity to laser phase noise.

In both clock and AI type proposals, two ensembles of atoms with a long-lived optically excited state are prepared in two satellites separated by a large distance.   Laser pulses transmitted between the two satellites interact with the atoms in order to imprint the effects of a passing gravitational wave onto atomic observables.  In both cases, the atoms are in a freely falling reference frame.  In AI proposals, this is accomplished by simply preparing the atoms in a high vacuum environment.  In clocks, the atoms are tightly trapped in an optical lattice formed by reflecting a laser off of a freely falling mirror that serves as an inertial reference, as in the proposed LISA interferometer \cite{LISApathfinder}. In AI proposals, the interaction between the atoms and the laser leads to recoil momentum kicks imparted to the atoms, while in clock proposals the photon momentum is absorbed by the much more massive inertial reference.  


In the clock community, gravitational wave detection has been described as a frequency measurement of Doppler shifts that result from the stretching of space between the satellites \cite{kolkowitz2016gravitational, vutha2015optical}.  
In the AI community, the stretching of space has been described as causing a phase shift between the laser phase and the atomic coherence \cite{graham2016resonant}.  
A key element of AI proposals is enhanced sensitivity to gravity waves through the use of Large Momentum Transfer (LMT) pulses in which the atoms acquire many photon recoil momentum kicks.  In contrast, the tight confinement of the atoms relative to the inertial reference mass in an optical lattice clock causes the photon recoil to be suppressed.  This difference in particular, as well as the language used to describe the devices, would seem to indicate that the two sensors are somehow  fundamentally different.

In sections II and III  we provide a comparison of these two sensors to show that the fundamental mechanism for sensitivity is in fact the same for the clock and AI type detectors. In both types of sensors, laser pulses that couple a ground and long-lived optically excited state imprint their local phase onto an internal quantum superposition state of each atom.  The atoms primarily act as a highly coherent phase memory that keeps track of these phase imprints and allows them to be read out through atomic observables.  



By viewing clock-like detectors as phase memories, rather than simply as clocks whose sole capability is to measure frequency, we show in section IV that they support the implementation of LMT-like protocols with enhanced sensitivity even when negligible momentum is transferred to the atom.  We further show in section V that the relevant phase information can actually be stored in stable ground states that evolve phases at acoustic rather than optical frequencies.  This enables useful coherent evolution times beyond the lifetime of the optically excited state.


\section{Optical Path Length Changes due to Gravitational Waves}
\begin{figure*}[htb]
\includegraphics[width=6.75in]{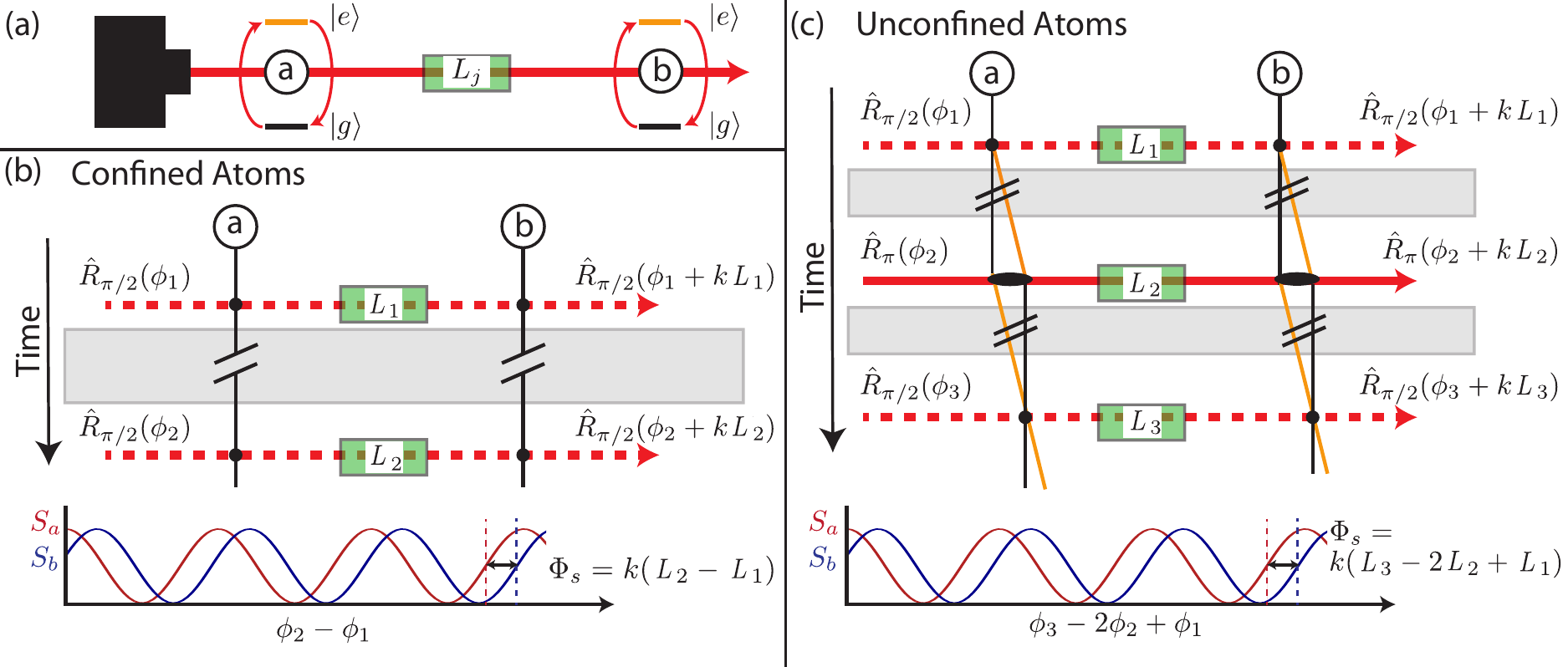}
\caption{A toy model for gravitational wave detection.  (a) Two atoms ($a$ and $b$) are addressed by a single laser resonant with a transition between the ground state $\ket{g}$ and a long-lived excited state $\ket{e}$.  A phase modulator mimics the effect of a gravitational wave by modifying the optical path length between the two atoms.  (b) For tightly confined atoms (clocks), a simple Ramsey sequence can be used to detect changes in optical path length.  A first $\pi/2$ pulse imprints the laser phase on the two atoms when the path length is shifted by $L_1$, and a second $\pi/2$ pulse acts when the path length is shifted by $L_2$.  Random variations in the laser phase ($\phi_1$ and $\phi_2$) are common to the two atoms, while phase shifts resulting from changes in path length are not.  The signal from the phase modulator manifests as phase shift $\Phi_s = k (L_2-L_1)$ between the excited state probability oscillations of the two atoms.  
(c) For unconfined atoms (atom interferometers), transitions between $\ket{g}$ and $\ket{e}$ are accompanied by momentum kicks, which necessitate an additional laser $\pi$ pulse in the middle of the sequence.  The phase shift between excited state oscillations of the two atoms is now $\Phi_s = k (L_3 - 2 L_2 + L_1)$.}
\label{fig:ToyModel}
\end{figure*}


For both types of proposals, the important effect of a gravitational wave is its modification of the optical path length $L$ between the two satellites.
We  will describe the effects of a gravitational wave at angular frequency $\omega_{g}$ by replacing the gravitational wave with a phase modulator that fills the space between the two satellites (e.g. an electro-optic modulator or EOM) sinusoidally driven at frequency $\omega_{g}$.
The drive applied to the modulator leads to a phase shift $\alpha(t)$ on laser light that is launched from one satellite and detected at the other satellite.   
The light is launched from the first satellite at time $t$, but arrives at the second satellite where it is detected at time $t+T_d$, where $T_d$ is the nominal delay or transit time between satellites. 

The total phase shift $\alpha(t_j)$ for a pulse launched at time $t_j$ is calculated by integrating along the optical path.  We can define a net effective path length for the pulse $L+L_j= L+\alpha(t_j)/k$ where the small change in optical path length is

\begin{equation}
L_j = \frac{h c}{2 \omega_{g}}\left(\sin{(\omega_{g} (t_j+T_d))} - \sin{(\omega_{g} t_j)}\right) \, .
\end{equation}

\noindent We have made the approximation that the gravitational wave's strain is very small $h\ll1.$  In this expression, $c$ is the speed of light in the undriven modulator, $k=2 \pi/\lambda$ is the wavenumber of the laser light, and $\lambda$ is the laser wavelength.  



\section{Detecting Changes in Optical Path Length}

The goal of this section is to understand how AI and clock-like sensors can be used to precisely estimate changes in the optical path length $L_j$. In our model, shown in Fig.~\ref{fig:ToyModel}, we consider two atoms labeled $a$ and $b$ that are separated by distance $L$. The atoms may either be cooled in free space (atom interferometers) or tightly confined (clocks).  The atoms have a long-lived optical transition, such as those in alkaline-earth and similar atoms (for example, Sr and Yb atoms).   A laser located near atom $a$ launches pulses of light that interact with both atoms with equal intensity.

The laser pulses interact with the atoms by applying so-called $\pi/2$ and $\pi$ pulses between ground $\ket{g}$ and excited $\ket{e}$ states.  We will assume that the coupling or Rabi frequency is much larger than the atomic decay rate from $\ket{e}$ and any relevant Doppler shifts due to atomic motion.  We will also assume for simplicity that the laser frequency is exactly equal to the atomic transition frequency.  These assumptions rule out some of the capabilities of atom interferometers (such as addressing atoms that have experienced different recoils independently), but retains the basic mechanism for sensitivity.  The effect of the laser pulses on the atoms written in a rotating frame at the atomic transition frequency can then be expressed using the operators: 
    


\begin{equation}
\hat{R}_{\pi/2} (\phi) = \frac{1}{\sqrt{2}}\begin{bmatrix}
    1       & -e^{-i (\phi + k x)} \\
    e^{i (\phi + k x)}       & 1 \\
\end{bmatrix}
\end{equation}

\begin{equation}\hat{R}_{\pi} (\phi) = \begin{bmatrix}
    0       & -e^{-i (\phi + k x)} \\
    e^{i (\phi + k x)}       & 0 \\
\end{bmatrix}
\end{equation}

\noindent These operators act on the basis states : 
\begin{equation}\ket{g}\otimes \Psi_g(x) = \begin{bmatrix}
    1       \\
    0       \\
\end{bmatrix} \otimes \Psi_g(x)
\end{equation}

\begin{equation}
\\ \ket{e}\otimes \Psi_e(x) = \begin{bmatrix}
    0       \\
    1       \\
\end{bmatrix} \otimes \Psi_e(x)
\end{equation}

\noindent that are a product of an internal state label $\ket{e}$ or $\ket{g}$, and an external state wavefunction $\Psi_{e,g}(x)$.  
Here $x$ represents the distance from a fixed plane where the laser's phase is defined as $\phi$.  The effect of these interactions is both to transfer amplitude between the internal states of the atoms and also to imprint the laser's local phase upon the transferred portion of the atom's wave function. The fact that $\Psi_g(x)$ may differ from $\Psi_e(x)$ indicates the possibility of entanglement between the internal and external degrees of freedom of the atom.  

For simplicity, we account for laser frequency noise by allowing the laser phases to vary between pulses, but taking the actual laser frequency to be fixed at the atomic transition frequency such that $k$ is constant \cite{RelaxedAssumption}.


First, we consider the effect of these rotations for a clock-like sensor in which the atoms are confined to much less than the laser wavelength (known as the Lamb Dicke regime.)  In this limit, we can think of the $e^{i(\phi + k x)}$ term as  imprinting a spatially constant phase onto the atom, whose value is determined by the location of the atom along the laser's path.  For example, a $\pi$ pulse applied to an atom tightly confined in a trap centered at position $x = A$ and in the ground state $\ket{\Psi_0}=\ket{g}\otimes \Psi_0(x)$ transfers the atom to the state $\ket{\Psi}\approx\ket{e}e^{- i(\phi + k A)}\otimes \Psi_0(x)$. The external state wave-function is to very good approximation  unmodified by the pulse, but its internal wavefunction has acquired a net phase $\phi+ k A$.



In contrast, in AI sensors the atoms are not confined and a change in the internal state is accompanied by a change in the external state. In particular, one cannot neglect the variation of the optical phase factor over the spatial extent of the atomic wavefunction. For concreteness, a $\pi$ pulse applied to an unconfined atom centered at $x=A$ and in the ground state  $\ket{\Psi_0}=\ket{g}\otimes \Psi_0(x)$ transfers the atom to the state $\ket{\Psi}=\ket{e}e^{- i\phi} \otimes e^{-i k x} \Psi_0(x)$.  As written, the internal portion of the wavefunction appears identical to that of the confined case, but now the external wavefunction has a spatially varying phase corresponding to one photon's worth of momentum recoil.


A change in the optical path length $L_j$  between the atom and the laser producing the $\pi$-pulse manifests in the confined case as $\ket{\Psi}= e^{- i(\phi + k(  L_j+A))}\ket{e}\otimes \Psi_0(x)$ and in the unconfined case as $\ket{\Psi}= e^{- i(\phi + k L_j)}\ket{e}\otimes \Psi_0(x)e^{- i k x}$.  The sensitivity to changes in path lengths from gravitational waves is due to the imprinting of an additional phase  $k L_j$, which is the same whether the atoms are confined or not.  In the case of AI sensors one must add additional pulses to become insensitive to the terms associated with the photon recoil.


\subsection{Clock-like Detectors}
A clock-like gravitational wave detector with confined atoms could be used to detect the gravitational wave phase shift using a basic Ramsey sequence \cite{kolkowitz2016gravitational}, pictured in Fig.~\ref{fig:ToyModel}b.  A change in the optical path length $L_j$ leads to a modification of the laser phase experienced by atom $b$ of $k L_j$.  The phase experienced by atom $a$ is unmodified by the change in optical path length.  In this sequence, the role of atom $a$ is then simply to record any variation of the phase of the laser itself due to technical sources of noise so that this laser  phase noise can be subtracted out from the final measurement.

Stepping through the Ramsey measurement, atoms in both locations are initially prepared in $\ket{g}$.  At time $t_1$, the laser drives the first $\pi/2$ pulse with phase $\phi=\phi_1$. we keep track of this phase only to demonstrate insensitivity to its value.  The rotation applied to atom $a$ is
$R_a = R_{\pi/2} (\phi_1)$.  When the same pulse arrives at atom $b$ it creates a rotation
$R_b = R_{\pi/2} (\phi_1+ k L_1)$.   
At a later time $t_2$, a second $\pi/2$ pulse is applied to atom $a$ with a laser phase $\phi=\phi_2$: $R_a = R_{\pi/2} (\phi_2)$. The same pulse of light travels to atom $b$ to drive a $\pi/2$ pulse
$R_b = R_{\pi/2} (\phi_2 + k L_2)$. 

The final signal that is detected is the difference in the probability for finding the atom in its excited and ground states. For atom $a$ this observable can be parameterized as  $S_a= P_{ea} -P_{ga} = \cos{\Phi_{as}}$ and for atom $b$ as  $S_b= P_{eb} -P_{gb} = \cos{\Phi_{bs}}$.  For the above Ramsey sequence, the signal phases are given $\Phi_{as}= (\phi_2 - \phi_1)$ and $\Phi_{bs}= k L_2- k L_1 + (\phi_2 - \phi_1)$,  The gravitational wave signal is given by extracting the difference of the phase of these two signals \cite{foster2002method}  

\begin{equation}
\Phi_s= \Phi_{bs}-\Phi_{as}= k(L_2 - L_1)
\end{equation}

\noindent The key result is that the technical laser phase noise  is canceled by having been recorded on both atoms and only the gravitational wave's signal remains.



\subsection{Atom Interferometer Detectors}
In an atom interferometer, the atoms are unconfined and a slightly more complicated sequence is needed.  The initial $\pi/2$ pulse entangles internal and external degrees of freedom of the atom by imparting a momentum kick to the portion of the atomic wavefunction transferred to $\ket{e}$.  If this momentum kick is not reversed, the portions of the atomic wavefunction will not be spatially overlapped to interfere at the time of the second $\pi/2$ pulse.  The simplest solution is to add a $\pi$ pulse in the middle of the sequence (Fig.~\ref{fig:ToyModel}c).  More complex sequences, with enhanced performance in different regimes are presented in \cite{graham2013new, graham2016resonant}, though the core mechanism for sensitivity is the same as this simple version.  

Stepping through the simplest AI sequence, the first $\pi/2$ pulse, $\pi$ pulse and second $\pi/2$ pulse are launched from the laser at times $t_1$, $t_2$, and $t_3$, respectively.  When the three pulses arrive at atom $b$ they will have experienced optical path length differences $L_1$, $L_2$, and $L_3$.

The signal that is extracted is the same as above, and depends only on the optical path lengths as 

\begin{equation}
\Phi_s = k (L_3-2 L_2 + L_1)
\end{equation}

\noindent where the random fluctuations in laser phase $\phi_3$, $\phi_2$, and $\phi_1$ are again cancelled because they are common to both atoms. A similar analysis for the atom interferometer is presented in \cite{yu2011gravitational}.

\section{Enhanced sensitivity analogous to large-momentum-transfer}

We now propose a mechanism by which the signal recorded by the clock-like sensor can be enhanced.  In atom interferometry in general, and AI based GW detectors in particular \cite{graham2013new}, large momentum transfer (LMT) is a crucial tool used to enhance the size of measured signals.  Our enhancement mechanism is very similar to LMT, but is applied to confined atoms, so no momentum is actually transferred to the atoms.  Instead, by allowing for multiple interactions between the lasers and atoms, a phase is repeatedly written in to the atomic coherence in a constructive manner to enhance the signal size $\Phi_s$.


\begin{figure}[!htb]
\includegraphics[width=3.375in]{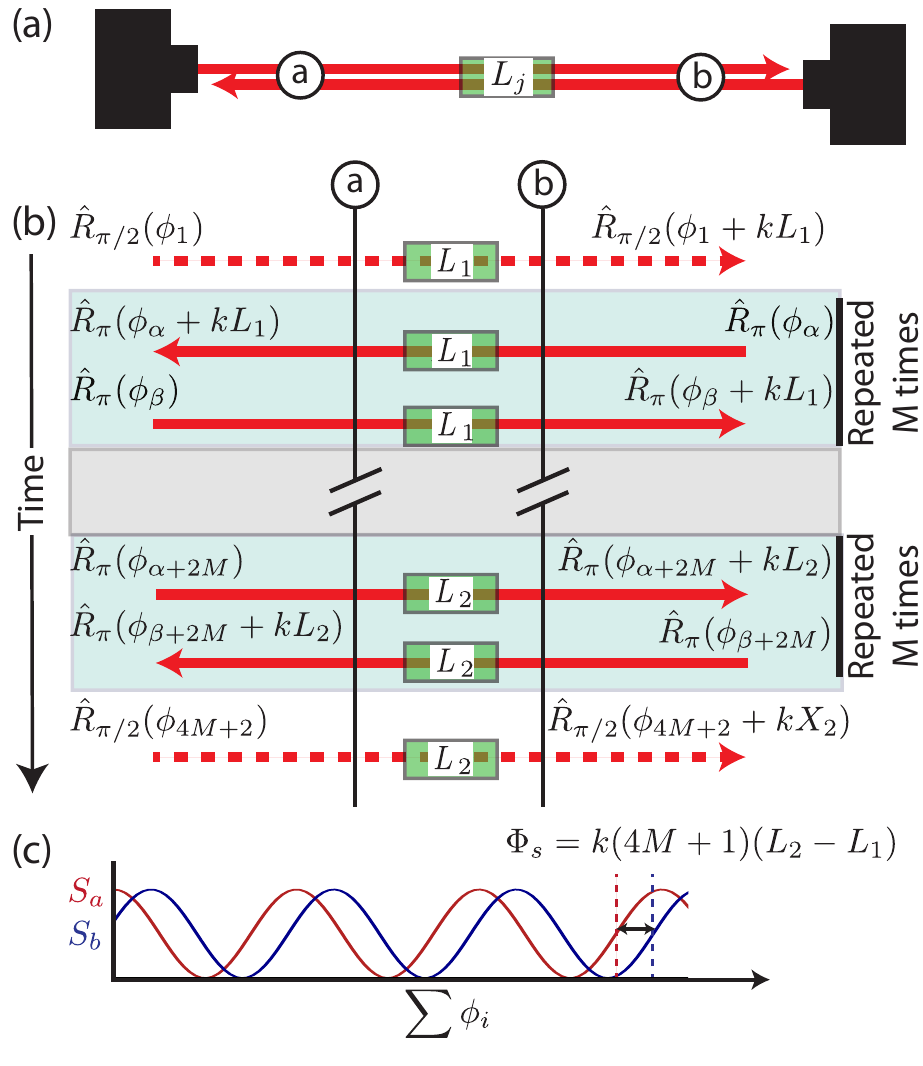}
\caption{LMT-like enhancement with confined atoms.  (a) Atoms are addressed by counter-propagating lasers, both of which pass through the phase modulator that sits between the atoms.  (b) Two blocks of $M$ pairs of $\pi$ pulses (blue boxes) are inserted between the $\pi/2$ pulses of the standard Ramsey sequence.  Each pair of $\pi$ pulses contains a pulse originating from each direction.  The duration of each block is assumed here to be short relative to the GW period, while a long evolution time (grey box) between the two blocks may be comparable to the GW period.  (c) The additional pulses lead to a factor of $4M+1$ enhancement in the signal phase $\Phi_s$ compared to simple Ramsey sequence.  }
\label{fig:LMT}
\end{figure}

The key to this enhancement sequence is to apply the laser pulses from alternating sides (as is proposed for LMT protocols for single-photon transitions with unconfined atoms \cite{graham2013new, hogan2016atom, graham2016resonant}) using one laser near atom $a$ and one laser near atom $b$, as illustrated in Fig.~\ref{fig:LMT}.  Because the sign of the laser phase shift imprinted on the atoms is opposite in sign when the atoms are driven from $\ket{e}$ to $\ket{g}$ versus $\ket{g}$ to $\ket{e}$, if the same laser were used to try to imprint its phase multiple times, it would simply unwrite the phase that it had just written in.  By interleaving $\pi$ pulses from a second laser, which does not encode the same gravitational wave phase shift as the first, we can ensure that the phase shift corresponding to the GW is always written in with the same sign.   The phases of the laser pulses are always referenced to a plane near the laser from which they originate so that there is negligible GW contribution to this phase.  


For simplicity, we treat the case where a set of pulses is applied in quick succession at the beginning of the measurement sequence, with launch times roughly equal to $t_1$ and with optical path length changes all equal to $L_1$.  During a subsequent free evolution period $T_e$, much longer than the time over which the rotations were applied, the optical path length may change.  At time $t_2=t_1+T_e$, a second set of rotations is quickly applied with launch times all roughly equal to $t_2$ such that all pulses experience an optical path length change equal to $L_2$ (a discussion of the effects of time delays is included in the appendix).  

Consider the first group of pulses. From the perspective of one of the two atoms, the laser pulse that arrives from the distant laser encodes in its phase $k L_1$ the change in path length.  This phase is then imprinted on the atomic superposition with a different sign between the portion that was transferred from $\ket{e}$ to $\ket{g}$ versus $\ket{g}$ to $\ket{e}$ such that the difference phase is $2 k L_1$.  The next laser pulse that is launched from the locally situated laser  resets the two portions of the atomic superposition to their original internal states. It imprints a phase shift of its own, but this phase does not encode information about the change in optical path length $L_1$, and the laser's phase noise cancels in the final differential signal phase $\Phi_s$.  Because the two portions of the atomic superposition have been reset to their original internal states, a subsequent rotation from the distant laser will imprint a phase encoding the same path length difference $k L_1$ that adds constructively with the previously imprinted phase so that the total imprinted phase on the atom is now $2 \times 2 k L_1 $.

If we apply $M$  pairs of $\pi$ rotations after the first $\pi/2$ pulse and $M$ pairs of $\pi$ rotations before the last $\pi/2$ pulse (i.e. $4M$ total launched $\pi$ pulses), with each pair containing a rotation originating from both the left and right sides, the differential phase shift between the two output channels is enhanced by a factor of $(4M+1)$ relative to the simple Ramsey sequence presented above so that now

\begin{equation}
\Phi_s= (4 M+1) k(L_2-L_1)
\end{equation}

\noindent In the regime we consider here, even a single pair of LMT pulses (i.e. $4$ total $\pi$-pulses), improves the estimate of $(L_2-L_1)$ by a factor of 25 in variance. As a result, the same precision can be achieved with a reduction in required resources such as atom number or averaging time by a factor of 25.   Alternatively, one could achieve the same sensitivity to gravitational waves with a 5 times shorter satellite separation or evolution time $T_e$ for reduced technical complexity or enhanced bandwidth for detecting gravitational waves, respectively.

This enhanced protocol is insensitive to laser phase noise for the same reason that the simple protocols presented in section III are:  each laser pulse interacts with both interferometers, so any phase noise on the laser cancels in the differential signal.  

Previous proposals for detectors using confined \cite{kolkowitz2016gravitational} and unconfined \cite{graham2016resonant} atoms include so-called dynamical decoupling (DD) sequences. A DD sequence would amount to applying a $\pi$ pulse from the same laser every time $t_j$ at which the magnitude of the optical path length change $\left|L_j\right|$ is maximal.  Because the sign of $L_j$ alternates between pulses, the resulting imprinted phase shifts add constructively.  This is conceptually similar to our LMT-like enhancement mechanism, except that we can switch the sign of the phase shift by alternating which laser applies the pulse instead of waiting for the sign of the GW to switch.  


Dynamical decoupling is useful when the evolution time $T_e$ greatly exceeds the GW period $T_g$.  In this regime, the enhancement in signal scales as $T_e/T_g \gg 1$. This reflects the fact that the phase shifts from $T_e/T_g$ cycles of the gravitational wave may be added constructively.  There is no constraint on delay time $T_d$. 

LMT-type sequences are useful when the period of the GW $T_g$ and total evolution time $T_e$ are long compared to the pulse transit time $T_d$.  This is necessary in order to send multiple pulses back and forth between the two satellites before the path length changes caused by the gravitational wave switches sign. The full signal response versus gravitational wave frequency is considered in the appendix. The main result is that one can build a long baseline experiment and set $T_e=T_d$ using a straightforward Ramsey sequence, or one can use LMT-type sequences to dramatically shorten the baseline such that $T_e$ is the same but now $T_e\gg T_d$.  In either case, the sensitivities are comparable. LMT-like enhancement allows one to address technical constraints that might be relaxed by operating with shorter baselines, at the expense of potentially introducing technical errors associated with the additional laser pulses.

Finally, as analyzed in \cite{graham2016resonant}, LMT can be combined with DD when both $T_e \gg T_g$ and $T_g \gg T_d$.  Doing so amounts to varying from which satellite the $\pi$ pulses are sent.

\section{Evolution times longer than the excited state lifetime}

We now present a further enhancement that can be realized by treating the atoms as a phase memory, rather than a clock.  In a typical clock, the atoms are considered to be a two-state system.  Because external degrees of freedom ideally remain unchanged, clocks lack additional quantum labels to specify other states.  Any pulse that interacts with one `arm' of the clock interferometer ($\ket{g}$) also interacts with the other ($\ket{e}$): the quantum mechanical amplitudes may be swapped between ground and excited states, but the portion of the wave-function in the ground state cannot be manipulated without also modifying the portion of the wave function in the excited state.  In an AI, the external degrees of freedom of the atom provide additional labels that allow the two arms to be manipulated independently.  

\begin{figure*}[!htb]
\includegraphics[width=6.5in]{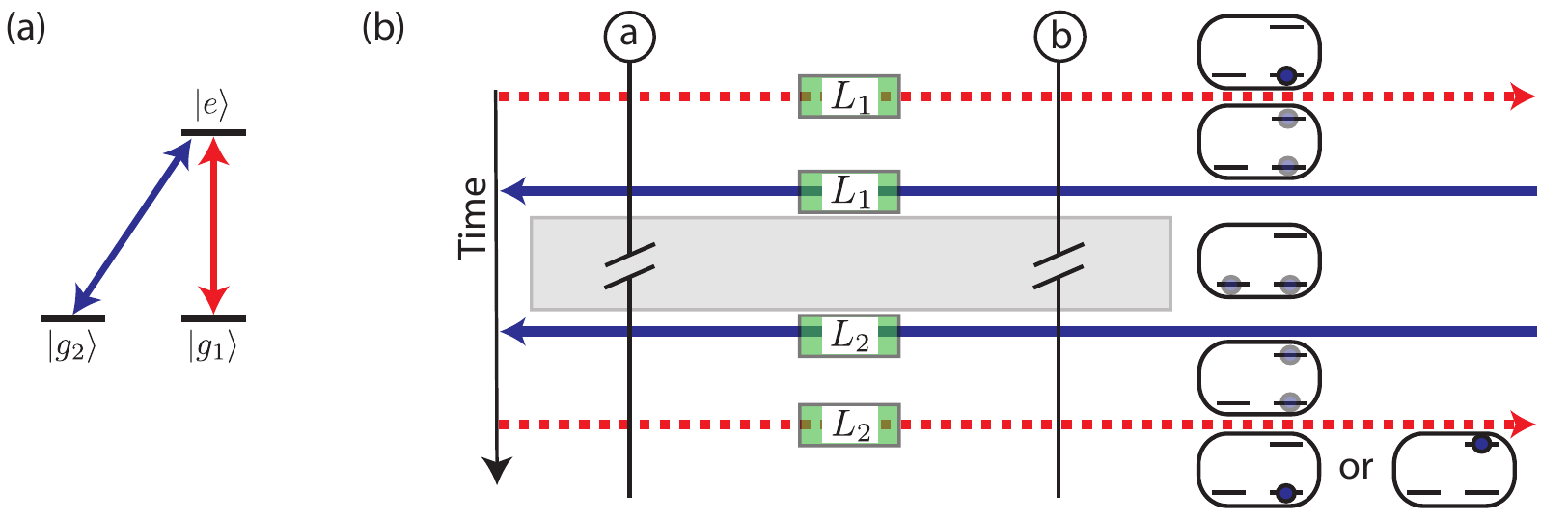}
\caption{Use of three-state system to allow atoms to stay in ground state for much of evolution time.  (a) Our procedure utilizes two ground states, $\ket{g_1}$ and $\ket{g_2}$ and a single long-lived optically excited state $\ket{e}$.  $\ket{g_1}$ and $\ket{g_2}$ can independently be coupled to $\ket{e}$ using laser pulses with different frequency and/or polarization (red and blue arrows). (b)  After preparing an atomic superposition of $\ket{g_1}$ and $\ket{e}$ using a $\pi/2$ pulse from the left (atoms begin in $\ket{g_1}$), the portion of the atoms in $\ket{e}$ is transferred to $\ket{g_2}$ using a $\pi$ pulse from the right.  The atoms are then in a superposition of $\ket{g_1}$ and $\ket{g_2}$ with phase dictated by the lasers and the path length $L_1$.  
The sequence of pulses is reversed after a long evolution time, during which time the atoms are not susceptible to excited state decay.  The phase shift in the excited state probabilities of the two atoms $\Phi_s = 2k(L_2-L_1)$ now encodes changes in the optical path length that occurred while the atoms were in the ground states.  }
\label{fig:3State}
\end{figure*}


While in clock type proposals, a fraction of the atomic wave-function must always be in the excited state to achieve sensitivity, AI type detection sequences may include useful periods where both arms of the interferometer are in the ground state \cite{graham2013new}, allowing for evolution times greater than the excited state lifetime.  This technique is particularly useful for enhancing the signal in the regime where the period of the GW exceeds the lifetime of the excited state.



In a recoil-free GW detector, additional internal states of the atoms can provide additional quantum labels that allow for the independent manipulation of only one arm of the interferometer.  Here, we consider an atomic species like $^{171}$Yb with nuclear spin $I=1/2$.  There are two ground states with nuclear spin projections $m_I=\pm 1/2$ that we we will label as $\ket{g_1}$ and $\ket{g_2}$.  For our purposes, we care about a single excited state $\ket{e}$, as shown in Fig.~\ref{fig:3State}a. A small magnetic field is applied to define a quantization axis. Transitions may be driven between either $\ket{g_1}$ and $\ket{e}$ or $\ket{g_2}$ and $\ket{e}$ by applying laser light with different polarizations or frequencies.  

An experimental sequence for such a protocol is shown in Fig.~\ref{fig:3State}b.  A superposition is prepared by driving a $\pi/2$ pulse originating from the left between $\ket{g_1}$ and $\ket{e}$.  A $\pi$ pulse originating from the right then transfers the population of $\ket{e}$ to $\ket{g_2}$.  The atoms remain in this configuration while the path length is allowed to change during an evolution time $T_e$, which we take to be much larger than the delay time $T_d$.  The pulse sequence is then reversed and the change in path length is read out via a population measurement. The resulting signal is $\Phi_s= 2k(L_2-L_1)$, again with no contribution from laser phase noise.  Memory storage in ground states could be combined with LMT-type enhancement by inserting $\pi$ pulses on the $\ket{g_1}$ to $\ket{e}$ transition in between the $\pi$ pulses on the $\ket{g_2}$ to $\ket{e}$ transition and the initial and final $\pi/2$ pulses to realize further enhancement.


Because the atoms are in superpositions of ground states, there is no spontaneous emission during $T_e$.  Instead, the relative phase that would normally  exist between $\ket{g_1}$ and $\ket{e}$ is now stored between $\ket{g_1}$ and $\ket{g_2}$ during the evolution period $T_e$.  In addition to extending the  evolution time beyond the spontaneous-decay limited coherence time of the atoms, this technique of shelving the atoms in ground states provides insensitivity to collisional effects that may limit the atomic density or coherence time achievable in practice.  Inelastic collisions between excited state atoms, which contribute to atom loss in dense systems, would be eliminated by storing the atoms in the ground states \cite{PhysRevLett.103.090801, bishof2011inelastic}.  Because elastic collisional properties of alkaline-earth-like atoms are independent of nuclear spin state \cite{Zhang19092014, scazza2014observation}, mean-field shifts due to elastic collisions are common to the two ground states and will not lead to noise or dephasing. Other bias errors in clocks such as black-body, electrostatic, and lattice polarizability and hyperpolarizability shifts would  also be suppressed.  As in the AI case, this technique is applicable when the evolution time $T_e$ would otherwise be limited to much less than the gravitational wave period $T_g$.

The key result is that it is not necessary for the phase memory to exist in a frame accumulating phase at an optical frequency.  It is only necessary that the interactions with the laser pulses happen between optical transitions, but these interactions can in principle represent a small fraction of the total evolution time.  

This protocol may appear similar to those that use two-photon Raman transitions in that the atoms occupy two ground states during a long evolution time.  However, Raman transitions require a laser pulse from each satellite to interact with an atomic ensemble at the same time in order to perform a rotation without populating the short-lived excited state.  
As with single-photon transitions, each laser pulse must also interact with both interferometers in order to obtain a signal that is insensitive to laser phase noise.  These two requirements lead to a conflict for detectors that use Raman transitions: to our knowledge, there is no protocol based on a two-satellite configuration in which each rotation is performed by simultaneously incident pulses from both satellites, and in which each pulse leads to the desired rotation in both interferometers. 
Because the use of single-photon transitions as discussed here allows population to be stored in the excited state for a time of order $T_d$ or longer, the laser pulses from the two satellites need not be simultaneously incident on the atoms, enabling phase-noise insensitive protocols that do not appear to be possible with Raman transitions.

\section{Conclusion}
Gravitational waves create phase shifts on optical pulses, which we would like to detect as sensitively as possible.  At the most basic level, one would like to store an optical pulse of light sent at time $t_1$ with phase shift $k L_1$ until a second pulse of light arrives with a different phase shift $k L_2$ at time $t_2$, then compare the phases of the two pulses.  Note that this is a related mechanism to that employed by purely optical GW detectors, which compare the phase shifts experienced by light travelling on two different paths that experience opposite GW-induced phase shifts, rather than light that that travelled along the same path at two different times.  
By providing a highly coherent phase memory with which the two pulses may interact, the atoms allow one to make time-separated phase comparisons of the two pulses in a manner that is insensitive to laser phase noise.  This capability eliminates the need in atomic detectors for a third satellite that is required in purely optical interferometers.

Further, multiple pulses can be made to constructively imprint their phases onto the atoms by properly alternating their launch direction (LMT) or by synchronizing their launch times with the frequency of the gravitational wave one wishes to detect (DD).  From a fundamental perspective, the ability to achieve LMT-type signal enhancement does not appear to require the transfer of momentum to the atoms or that the two lasers originate from different directions (although the latter is required for cancellation of technical sources of noise).  In certain regimes, LMT allows the signal size to be increased such that the uncertainty in the estimate of the optical phase of interest  $k(L_2-L_1)$ can be greatly reduced to well below the atomic standard quantum limit  $1/\sqrt{N}$~rad on atomic phase resolution.  

For gravitational wave detection, we do not need to measure the absolute frequency of a laser relative to an atomic transition frequency, as one would do in a clock. As we have shown, this allows for the construction of a ground state shelving protocol with reduced sensitivities to perturbations and evolution times $T_e$ greater than the optical transition lifetime.

Ultimately, the decision to use confined or unconfined atoms will depend on a myriad of technical considerations that we have no business weighing in on.  From a fundamental perspective, both methods have the same mechanism for sensitivity: atom interferometers and clocks both sense changes in phase that result from changes in optical path length between the two satellites. 
This interpretation should help distinguish to what degree differing expected sensitivities for future proposals are due to the choice of sensor architecture versus the specific parameters considered.




\section{ACKNOWLEDGMENTS}\label{ack}
We thank Shimon Kolkowitz, Jason Hogan and Peter Graham for important discussions and input.  
All authors acknowledge financial support from DARPA QuASAR, ARO, NSF PFC, and NIST. J.R.K.C. acknowledges financial support from NSF GRFP. This work is supported by the National Science Foundation under Grant Number 1125844.

\section{Appendix: Accounting for Time Delays}

In the main text, we considered an LMT pulse sequence in which the time delay $T_d$ is much shorter than both the evolution time $T_e$ and gravitational wave period $T_g$.  Rather than launching a series of LMT pulses at the beginning and at the end, one could in principle launch a continuous series of alternating $\pi$-pulses during the entire period of time between the $\pi/2$ pulses.  For continuous LMT, we find that the signal size averaged over all phases of the gravitational wave is (to good approximation) given by:

\begin{equation}
\bar{\Phi}_s = 4 \sqrt{2} h \frac{\omega_l \left|\sin(\omega_g T_d /2) \right|}{ \omega_g^2 T_d} \sin^2( \omega_g T_e /4)
\end{equation}

\noindent where $\omega_l$ is the laser angular frequency.  When $T_d\gg T_g$, the sensitivity falls of rapidly as $1/\omega_g^2$.   However, in the limit $T_d\ll T_g $ the scaling changes to

\begin{equation}
\bar{\Phi}_s \approx 2 \sqrt{2} h \frac{\omega_l}{\omega_g} \sin^2( \omega_g T_e /4)
\end{equation}

\noindent  For comparison, if the spacing of the satellites is increased until $T_d = T_e$, then one cannot use LMT and the Ramsey sequence yields an averaged signal:

\begin{equation}
\bar{\Phi}_s =\sqrt{2} h \frac{\omega_l}{\omega_g} \sin^2( \omega_g T_e /2)
\end{equation}

\noindent The oscillations in the signal versus $\omega_g$ differ by a factor of 2 and the envelope of the signal size is larger by a factor of 2 for LMT.  Most importantly, approximately the same signal can be obtained using continuous LMT with of order $T_e/T_d$ $\pi$ pulses and an approximately $T_e/T_d$ times shorter satellite satellite spacing. It is likely that decreasing the spacing of the satellites will be advantageous for technical reasons, for instance, it allows for reduced requirements on laser power and pointing stability.

\bibliography{ThompsonLab.bib}
\end{document}